\documentclass[printer]{aa}

\usepackage{graphicx}
\usepackage{natbib}
\usepackage{amssymb}
\usepackage{amsmath}
\usepackage{xspace}
\usepackage[dvipsnames]{xcolor}
\usepackage{hyperref}
\usepackage{dblfloatfix}

\usepackage[normalem]{ulem}
\usepackage[varg]{txfonts}

\newcommand{\rb}[1]{{\color{black}  #1}} 
\newcommand{\mpo}[1]{{\color{black}  #1}} 

\graphicspath{{./}{figures/}}

\begin{document}

\title{Morphology of \rb{supernova remnants} and their halos}

\author{R. Brose \inst{1,2}\fnmsep\thanks{Corresponding author, \email{robert.brose@desy.de}} \and
    M. Pohl  \inst{1,3}\and
    I. Sushch  \inst{4,5}}

\institute{Institute of Physics and Astronomy, University of Potsdam, 14476 Potsdam, Germany 
\and Dublin Institute for Advanced Studies, Astronomy \& Astrophysics Section, 31 Fitzwilliam Place, D02 XF86 Dublin 2, Ireland
\and DESY, 15738 Zeuthen, Germany
\and Centre for Space Research, North-West University, 2520 Potchefstroom, South Africa
\and Astronomical Observatory of Ivan Franko National University of Lviv, Kyryla i Methodia 8, 79005 Lviv, Ukraine
}
\date{Received ; accepted}



\abstract
{Supernova remnants are known to accelerate particles to relativistic energies, on account of their non-thermal emission. The observational progress from radio to gamma-ray observations reveals more and more morphological features that need to be accounted for when modeling the emission from those objects.}
{We use our time-dependent acceleration code RATPaC to study the formation of extended gamma-ray halos around supernova remnants and the morphological implications that arise when the high-energetic particles start to escape from the remnant.}
{We performed spherically symmetric 1-D simulations in which we simultaneously solve the transport equations for cosmic rays, magnetic turbulence, and the hydrodynamical flow of the thermal plasma in a volume large enough to keep all cosmic rays in the simulation. The transport equations for cosmic-rays and magnetic turbulence are coupled via the cosmic-ray gradient and the spatial diffusion coefficient of the cosmic rays, while the cosmic-ray feedback onto the shock structure can be ignored. Our simulations span 25,000 years, thus covering the free-expansion and the Sedov-Taylor phase of the remnant’s evolution.}
{We find strong difference in the morphology of the gamma-ray emission from supernova remnants at later stages dependent on the emission process. At early times both - the inverse-Compton and the Pion-decay morphology - are shell-like. However, as soon as the maximum-energy of the freshly accelerated particles starts to fall, the inverse-Compton morphology starts to become center-filled whereas the Pion-decay morphology keeps its shell-like structure. Escaping high-energy electrons start to form an emission halo around the remnant at this time. There are good prospects for detecting this spectrally hard emission with the future Cerenkov Telescope Array, as there are for detecting variations of the gamma-ray spectral index across the interior of the remnant. Further, we find a constantly decreasing non-thermal X-ray flux that makes a detection of X-ray unlikely after the first few thousand years of the remnants evolution. The radio flux is increasing throughout the SNR's lifetime and changes from a shell-like to a more center-filled morphology later on.
}
{}
 

\maketitle


\section{Introduction}

The detection of extended gamma-ray emission around two nearby pulsars - Geminga and PSR B0656+14 \citep{2017Sci...358..911A, 2017ApJ...843...40A} - spawned an extensive discussion of TeV-halos around pulsars \citep{2020A&A...636A.113G}.

Escaping high-energy electrons and positrons can be confined close the accelerating pulsar-wind nebulae (PWN) by self-amplified magnetic turbulence and enhance the TeV gamma-ray emission in the vicinity of the PWN \citep{2018PhRvD..98f3017E}. The resulting halos will have an energy-dependent morphology, with a smaller extension towards the highest energies \citep{2020arXiv200611177P}.  

However, the interpretation of the observational data is not trivial. Simple uniform-diffusion models can explain the observed morphology of the gamma-ray emission on account of confining all electrons and positrons close to the pulsar. A contribution to the positron excess observed in the AMS-data \citep{2008Natur.456..362C, 2009Natur.458..607A} would thus be ruled out. In reality, a spatially non-uniform diffusion coefficient has to be expected for self-amplified turbulence and could both explain the observed morphology and allow for sufficient escape flux to support the local positron-flux \citep{2018PhRvD..97l3008P}.   

A similar self-regulation of the diffusion coefficient by escaping particles is known to exist around supernova remnants (SNRs) as well. There, the scattering turbulence is created mainly by escaping hadrons, and the electrons accelerated at the SNR blast-wave are trapped as a side effect \citep{2010A&A...513A..17O, 2016MNRAS.461.3552N}.

So far, there has not been a direct measurement of \rb{cosmic rays (CRs)} escaping from a SNR. A possible scenario would be CRs illuminating molecular clouds close the SNRs where the enhanced target density boosts hadronic gamma-ray emission \citep{2012ApJ...745..140Y}. There is evidence of larger extension of the gamma-ray emission around RXJ1713.7-3946 compared to the X-ray emission that could indicate CR escape \citep{2018A&A...612A...6H}. However, there is indirect evidence that CR escape has to happen around SNRs. Recent studies of particle acceleration in supernova remnants showed that typically soft, broken power-law spectra of aged SNRs \citep{2019ApJ...874...50Z} can be produced by the escape of particles at the highest energy from the interior of the SNR. Here, the most energetic particles escape once the SNR is not capable of accelerating them any further, creating soft spectra inside the SNR where most of the emission is produced \citep{2020A&A...634A..59B,2019MNRAS.490.4317C}. 

These concepts potentially resolve the tension between \rb{the $s=2$ spectra predicted by} shock-acceleration theory and the \rb{typically soft emission} spectra \rb{of $s\geq2.7$} observed inside \rb{evolved} SNRs\rb{, especially in the gamma-ray domain}. However, the question whether SNRs are the sources of the Galactic CRs is yet still unanswered. The earlier studies show that there is no \rb{contradiction} between the observed soft emission spectra and the somewhat harder total-production spectra \rb{with $s\approx2.2-2.4$ that are predicted for the sources of Galactic CRs by Galactic propagation models \rb{\citep{2011ApJ...729..106T}}. The \mpo{accumulated proton spectrum is} approximately} the \mpo{same as the time integral} over the SNRs lifetime and represents the CR yield that gets finally released into the sea of Galactic CRs once the SNR-shock faded.

The aim of this paper is to explore the observational signatures that can be expected from escaping CRs around SNRs.

\section{Basic equations and assumptions}
The methodology is similar to that of \cite{2020A&A...634A..59B}, and here we shall only give a short summary. We combine a kinetic treatment of the CRs with a thermal leakage injection model, a fully time-dependent treatment of the magnetic turbulence, and a PLUTO-based simulation of the hydrodynamical flow profiles.

\subsection{Cosmic rays}
We solve the kinetic equation for the differential number density of CRs, $N$,
\begin{align}
\frac{\partial N}{\partial t} =& \nabla(D_r\nabla N-\mathbf{u} N)\nonumber\\
 &-\frac{\partial}{\partial p}\left( (N\dot{p})-\frac{\nabla \cdot \mathbf{ u}}{3}Np\right)+Q
\label{eq:CRTE}
\end{align}
in the test-particle limit, where $D_r$ denotes the spatial diffusion coefficient, $\textbf{u}$ the advective velocity, $\dot{p}$ energy losses \rb{(see section \ref{sec:losses})}, and $Q$ the source of thermal particles \citep{Skilling.1975a}.

We rewrite equation (\ref{eq:CRTE}) by transforming the radial coordinate, $r$, to a new spatial coordinate that is co-moving with the SNR shock and provides a very high numerical resolution at the shock front \citep[and references therin]{2020A&A...634A..59B}.
The sea of Galactic CRs is neglected in this simulations. As long as the fraction of injected particles is constant over time, the emission from freshly injected particles is always dominant over the contribution from background CRs.

\subsubsection{Injection}
We use a thermal leakage model \citep{Blasi.2005a,1998PhRvE..58.4911M} for the injection of particles. Here, the efficiency of injection $\eta_i$ is given by
\begin{align}
\eta_i = \frac{4}{3\sqrt{\pi}}(\sigma-1)\psi^3e^{-\psi^2}\text{ , }\label{eq:Injection}
\end{align}
where $\sigma$ is the shock compression ratio, and $\psi$ is the multiple of the thermal momentum, at which we inject particles.

\rb{Several authors noted that the bipolar morphology in the non-thermal emission of SN 1006 can be attributed to effects of the shock-obliquity on the injection or acceleration efficiency \citep{2003A&A...409..563V, 2009MNRAS.395.1467P, 2012MNRAS.419.1421B, 2020MNRAS.498.5557P}. This \mpo{notion} seem to be supported by the hybrid-simulations of \cite{2014ApJ...783...91C}, where the acceleration at quasi-perpendicular shocks is strongly suppressed. \cite{2013MNRAS.430.2873R} on the other hand found a quasi-universal behaviour of shocks irrespective of the magnetic-field orientation very far upstream of the shocks by using a spherical-harmonics expansion of the CR Fokker-Planck equation. This suggests that injection may only be weakly dependent on the shock orientation. We have to ignore possible obliquity effects in our spherically symmetric model and assume a quasi-parallel configuration across the shock-surface.}

This injection scenario is a simplification, in particular for electrons, for which pre-acceleration to a few tens of MeV is required and established at the shock \citep{Matsumoto2017,2018PhPl...25h2103L,2019ApJ...878....5B}. We are interested in particles at energies well above $100$~MeV, and so the particulars of that pre-acceleration can be ignored.

The multiple of the thermal momentum, $\psi$, determines the fraction of thermal particles that get turned into CRs. However, $\psi$ and hence $\eta$ are only weakly constrained. Low values for $\psi$ are required in scenarios featuring non-linear modifications of the shock structure by CR pressure. The originally proposed lower boundary  $\psi\approx3.5$ \mpo{ensures that} thermal particles \rb{have a mean-free path larger than the shock-thickness and \mpo{can} participate in the DSA-process}. For this work we choose $\psi=4.2$, which guarantees a CR pressure of less than $2.5\,$\% of the shock ram pressure during the entire simulation. Likewise, this value is close to the amount of injected particles seen in SN 1006, where the likely leptonic origin of the emission allows a reasonably sound estimate of $\eta_i$  (see section \ref{sec:InjSN1006} for details). We have chosen an equal amount of injected electrons and protons. 

\subsubsection{Inverse-Compton losses}\label{sec:losses}
In addition to the synchrotron losses \rb{that were considered already in earlier versions of RATPaC}, electrons will also suffer losses by collisions with photons from background photon-fields. Usually these inverse-Compton (IC) losses can be neglected as synchrotron losses in amplified magnetic field will dominate. However, high-energy electrons spend a large fraction of time upstream of the shock, where IC and synchrotron losses are of similar strength.

We used the approximations derived by \cite{2006ApJ...644.1118R} to account for IC losses by collisions with the cosmic microwave background. We note that additional photon fields might have to be taken into account when modeling core-collapse supernova remnants where strong local infrared and optical photon fields have to be expected or when specific Type-Ia SNRs are modeled for which estimates of the background photon-fields are available based on their location in the Galactic disk. 

\subsection{Magnetic turbulence}
In parallel to the transport equation for CRs, we solve a transport equation for the magnetic-turbulence spectrum, assuming Alfv\'en waves only. The temporal and spatial evolution of the spectral energy-density per unit logarithmic bandwidth, $E_w$, is described by
\begin{align}
 \frac{\partial E_w}{\partial t} +  \cdot \nabla (\mathbf{u} E_w) + k\frac{\partial}{\partial k}\left( {k^2} D_k \frac{\partial}{\partial k} \frac{E_w}{k^3}\right) = \nonumber\\
=2(\Gamma_g-\Gamma_d)E_w \text{ . }
\label{eq:Turb_1}
\end{align}
Here, $\mathbf{u}$ denotes the advection velocity, $k$ the wavenumber, $D_k$ the diffusion coefficient in wavenumber space, and $\Gamma_g$ and $\Gamma_d$ the growth and damping terms, respectively \citep{2016A&A...593A..20B}.

We calculate the diffusion coefficient from $E_w$ using 
\begin{align}
    D_{r} &= \frac{4 v}{3 \pi }r_g \frac{U_m}{{E}_w} \text{ , }
\end{align}
where $U_m$ denotes the energy density of the large-scale magnetic field, $v$ is the particle velocity, and $r_g$ the gyro-radius of the particle.

As initial condition, we used a diffusion coefficient, and hence a turbulence spectrum, as suggested by Galactic propagation modeling \citep{2011ApJ...729..106T}, but reduced by a factor ten on account of numerical constraints, 
\begin{align}
    D_0 &= 10^{28}\left(\frac{pc}{10\,\text{GeV}}\right)^{1/3}\left(\frac{B_0}{3\,\mu\text{G}}\right)^{-1/3} \text{ . }
\end{align}
This choice is a factor of ten higher then in our previous works.

We use growth-rate based on the resonant streaming instability \citep{Skilling.1975a, 1978MNRAS.182..147B},
\begin{align}
    \Gamma_g &= A\cdot\frac{v_\text{A}p^2v}{3E_\text{w}}\left|\frac{\partial N}{\partial r}\right| \text{ , }
\end{align}
where $v_\text{A}$ is the Alfven-velocity. We introduced a linear scaling factor, $A$, to artificially enhance the amplification. We used $A=10$ throughout this paper to mimic the more efficient amplification due to the non-resonant streaming instability \citep{2000MNRAS.314...65L, 2004MNRAS.353..550B}. The particulars of the non-resonant amplification are beyond the present capabilities of our code since the back-reaction of CR streaming that terminates the wave-growth, a modification of the bulk flow \citep{2009ApJ...694..626R, 2010ApJ...709.1148N,2017MNRAS.469.4985K}, can not be accounted for. Likewise difficult to handle, and in fact quite unclear, is the scattering efficiency of the non-resonant modes. 
The value we choose for $A$ guarantees a cut-off energy in the gamma-ray spectrum between $1-10\,$TeV as observed in young SNRs. 

We calculate the total magnetic-field strength as 
\begin{align}
    B_\text{tot} &= \sqrt{B_0^2+B_\text{Turb}^2} \text{ , }
\end{align}
where $B_0$ is the large-scale magnetic field. We solve the induction equation to model the transport of the frozen-in large-scale magnetic field \citep{2013A&A...552A.102T}. The far-upstream field is assumed to be uniform with strength $5\mu$G \rb{and the field is assumed to be fully turbulent, resulting in a magnetic field-compression of $B_\text{d}=\sqrt{11}B_\text{u}$}.

Since we exceed the growth rate of the resonant streaming instability \citep{1978MNRAS.182..147B} by a factor of ten, the turbulent field is amplified to $\delta B\gg B_0$ during the initial phases of SNR evolution. The peak amplitude of the field is reached right at the shock. Downstream the field strength quickly falls due to efficient cascading of turbulence. The resulting magnetic-field profiles resemble the profiles suggested by \cite{2005ApJ...626L.101P}. In the early phases of SNR evolution peak amplitudes of $\approx100\,\mu$G are reached, whereas after $1,000\,$years one finds $\approx40\,\mu$G - a value compatible with the $30\,\mu$G estimated for the $1,000\,$year-old SNR SN1006 \citep{2010A&A...516A..62A}. 

The growth of the magnetic turbulence and hence the magnetic field is balanced by cascading. This process is described as a diffusion process in wavenumber space, and the diffusion-coefficient is given by \citep{1990JGR....9514881Z, Schlickeiser.2002a}
\begin{align}
    D_\text{k} &= k^3v_\text{A}\sqrt{\frac{E_\text{w}}{2B_0^2}} \text{ . }
\end{align}
This phenomenological treatment will result in a Kolmogorov-like spectrum, if cascading is dominant. Since $v_\text{A}\propto B_\text{tot}$, the cascading rate will depend on the level of magnetic turbulence in two different regimes
\begin{align}
    D_k \propto 
    \begin{cases}
        \sqrt{E_\text{w}} &\text{for $E_\text{w} < B_0^2/8\pi$}\\
        E_\text{w}^{3/2} &\text{for $E_\text{w} > B_0^2/8\pi$} \text{ . }\\ 
    \end{cases}
\end{align}
Once the turbulent field dominates over the background field, the cascading rate depends more sensitively on the energy density of magnetic turbulence. Without requiring other damping mechanisms, the enhanced cascading efficiently limits the maximum level of turbulence to a level commensurate with that derived from SNR observations and PIC simulations \citep{2006ESASP.604..319V, 2009ApJ...694..626R, 2010ApJ...709.1148N}. 


\subsection{Thermal plasma}
In the test-particle limit, the evolution of an SNR can be described with the standard gas-dynamical equations:
\begin{align}
\frac{\partial }{\partial t}\left( \begin{array}{c}
                                    \rho\\
                                    \textbf{m}\\
                                    E
                                   \end{array}
 \right) + \nabla\left( \begin{array}{c}
                   \rho\textbf{v}\\
                   \textbf{mv} + P\textbf{I}\\
                   (E+p)\textbf{v} 
                  \end{array}
 \right)^T &= \left(\begin{array}{c}
                    0\\
                    0\\
                    L
                   \end{array}
 \right)\\
 \frac{\rho\textbf{v}^2}{2}+\frac{P}{\gamma-1}  &= E \text{,}
\end{align}
where $\rho$ is the density of the thermal gas, $\textbf{v}$ the plasma velocity, $\textbf{m}=\textbf{v}\rho$ the momentum density, $P$ the thermal pressure of the gas, $L$ the energy losses due to cooling, and $E$ the total energy density of the ideal gas with $\gamma=5/3$. We solve this system of equations under the assumption of spherical symmetry in 1-D using the PLUTO code \citep{2007ApJS..170..228M}. The non-equilibrium cooling function, $L$, is taken from \cite{1993ApJS...88..253S}.

In this work, we display results for type-Ia supernova explosions. We initiate the simulations with exponential-ejecta profiles:
\begin{align}
 \rho_\mathrm{SN} =& A\exp(-v/v_e)t_i^{-3,}\text{ and } v = r/t_i, \label{eq:Vik_1} \\
 \text{ with } & v_e = \left( \frac{E_\mathrm{ex}}{6M_\mathrm{ej}} \right)^{1/2,} \text{ and } A =\frac{6^{3/2}}{8\pi}\frac{M_\mathrm{ej}^{5/2}}{E_\mathrm{ex}^{3/2}}\label{eq:Vik_2}
\end{align}
as initial conditions \citep{1998ApJ...497..807D}. Here,  $t_i=2.5\,\mathrm{yrs}$ is the start time of our simulation, $M_\mathrm{ej}=1.4M_\mathrm{sol}$ the ejecta mass, $E_\mathrm{ex}=10^{51}\,$erg the explosion energy, and $r$ the spatial coordinate. The density of the ambient medium was chosen to be $0.4\,\text{cm}^{-3}$.




\section{Results}

We followed the evolution of the remnant for $25,000\,$years. The remnant enters the Sedov-Taylor phase after $1,300\,$years\footnote{The use the initial conditions Eq. (\ref{eq:Vik_1}-\ref{eq:Vik_2}) results in a continous, smooth variation of the expansion parameter $m=V_\textbf{sh}/(R_\text{sh}\cdot t)$. After 1,300 yrs $m=0.47$, reasonabliy close to the asymptotic $m=0.4$ clasically expected for the Sedov-phase.} and would enter the post-adiabatic phase after $35,000\,$years.

The total magnetic field reaches $90\,\mu$G after $300\,$yrs and drops to $40\,\mu$G after $1000\,$yrs. After $10,000\,$yrs, the turbulent field is weaker than the compressed large-scale field.

In the following, we first describe the morphology of the TeV-halo over the lifetime of the remnant and then examine its detectability with today's and future gamma-ray experiments. 

\subsection{Halo-evolution}

\begin{figure}[ht]
\includegraphics[width=0.485\textwidth]{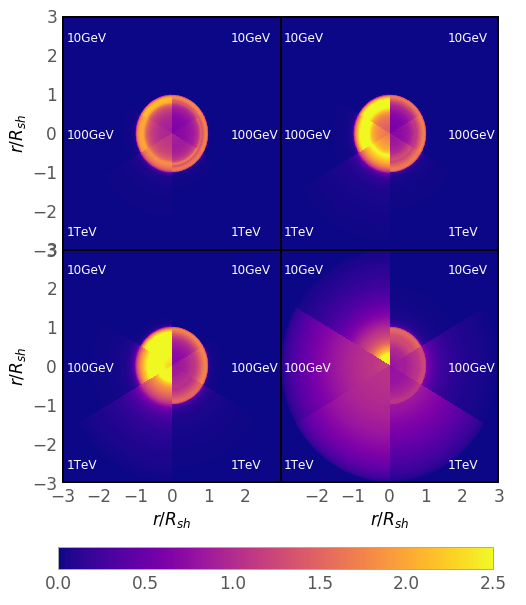}
\caption{Maps of normalized surface brightness of a Type-Ia SNR after 300, 1000, 2,000 and 10,000 yrs (from top left to bottom right). Left hemispheres are for IC emission and right hemispheres for PD emission.} 
\label{fig:EmissionMaps}
\end{figure}

We calculated intensity maps for \rb{inverse-Compton and Pion-decay (PD)} emission at three energies over the lifetime of the SNR. The results are presented in Figure \ref{fig:EmissionMaps}.

Initially, both radiation mechanisms produce a shell-like morphology in all energy bands. From roughly $1000\,$years on, the shell thickness of IC emission exceeds that of PD radiation, because the latter is boosted by the high gas density immediately downstream of the shock. Contrary, the IC emission reflects only the distribution of electrons, and already after $1000\,$years we notice IC emission outside the SNRs shell. The age of $1000\,$years roughly corresponds to the time when the SNR luminosity peaks regardless of the \rb{gamma-ray} emission mechanism, on account of the transition to the Sedov-Taylor stage. For an ambient density of $0.4\,$cm$^{-3}$ and an average free-expansion velocity of $5000\,$km/s, five solar masses of material will have been swept-up by the shock within the first 1000 years of expansion\footnote{The initial shock-velocity might exceed $10,000\,$ km/s but it is not constant even for the free-expansion phase.}. The transition to the Sedov-phase usually also marks the time of the highest maximum energy of particles \citep{2003A&A...403....1P}. At later times, particles of the highest energy start to escape from the remnant \citep{2020A&A...634A..59B}. 

Even after $10,000\,$years, the PD morphology remains shell-like whereas that of IC emission becomes center-filled. Even low-energy CRs propagated into the center of the remnant, and the projection enhances the brightness towards the center of the remnant in the IC channel. The low density of the thermal plasma in the center keeps suppressing PD emission from the central region. As more and more high-energy electrons escape with increasing age, an extensive halo of $>100$-GeV electrons is formed around the remnant. There is also a faint halo in the PD channel after $10,000\,$years, but it is much weaker than the IC halo on account of the low gas density. 

Synchrotron losses most strongly modify the distribution of high-energy electrons. They cause the thinner IC shell at $1\,$TeV compared to lower energies that is visible in the intensity maps. The magnetic field peaks at the shock, and so synchrotron losses are strongest there. The same effect is responsible for the relative smaller extend of the bright IC-region at $2,000\,$years. In the unshocked ejecta the field strength is $0.2\,\mu$G, making IC scattering the dominant energy-loss mechanism. Generally, the IC-loss timescale exceeds that for synchrotron losses where the magnetic field surpasses $5\,\mu$G.    

\begin{figure}[h]
\includegraphics[width=0.485\textwidth]{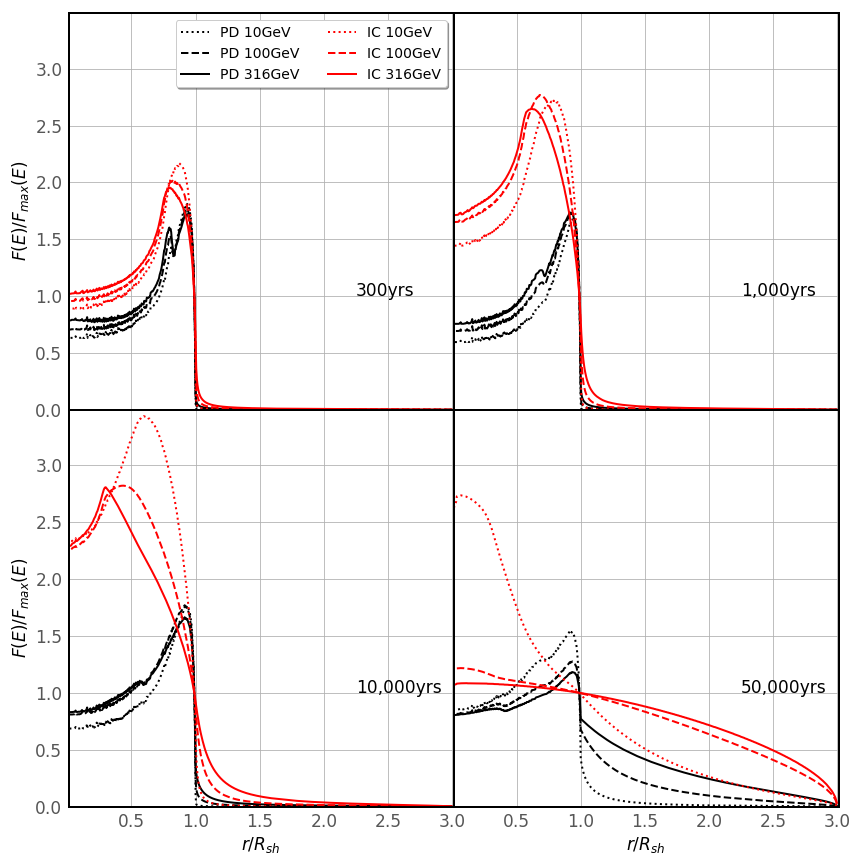}
\caption{Projected intensity profiles normalized to the surface brightness at the shock at four stages of SNR evolution.} 
\label{fig:EmissionProfiles}
\end{figure}

Figure \ref{fig:EmissionProfiles} shows the projected profiles of gamma-ray intensity at four stages of the remnant evolution. 
It is clearly visible that the extension of the IC halo increases with energy, in contrast to the case of PWNs. The size of the halo is not defined by a balance between acceleration of the highest energetic electrons and their synchrotron cooling, as all high-energetic particles have been accelerated at early times. More important is that the high-energy particles experience the largest diffusion coefficient and thus can fill a larger halo. 

\begin{figure*}[hbt]
\includegraphics[width=0.99\textwidth]{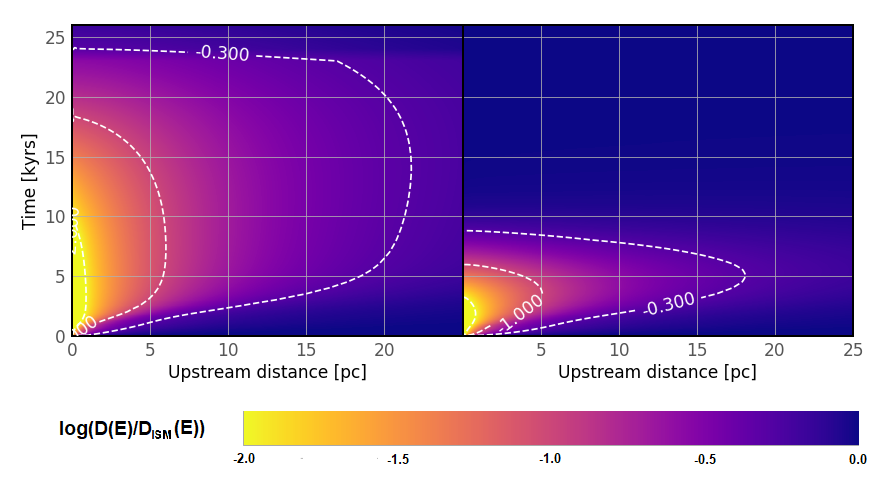}
\caption{Normalized diffusion coefficient in the upstream region as function of distance to the shock and the SNR age. The left panel is for a particle energy of $100\,$GeV, the right panel for $3\,$TeV. The dashed lines correspond to $1\,$\%, $10\,$\% and $50\,$\% of $D_\text{ISM}$.} 
\label{fig:Diffusion}
\end{figure*}

\rb{The morphology of the remnant also strongly depends on irregularities in the ambient medium. Deviations from spherical symmetry may be caused if the coherence length of the turbulent ambient magnetic field is larger than the size of the remnant \citep{2020MNRAS.498.5557P}. Similar effects have been described earlier for the hadronic and leptonic emission morphology of remnants expanding in a uniform ambient field \citep{2009MNRAS.395.1467P, 2012MNRAS.419.1421B}. Our results qualitatively agree with \mpo{these conclusions} in regions where the field is parallel to the shock or where the coherence-scale of the ambient field is smaller than the size of the remnant. However, accelerated diffusion into the center of the remnant with the onset of the Sedov-stage produces a more center-filled morphology for the IC-emission than can be obtained by models relying solely on advection.}

A comparison of our predictions with measurements is difficult as there is no firm detection of an emission halo around a SNR (see also section \ref{sec:detectability} for details). However, extended gamma-ray emission around the two Pulsars Geminga and PSR  B0656+14 has been detected \citep{2017Sci...358..911A}. The measurements show a roughly exponential decrease of the surface-brightness with increasing radial distance from the central pulsar. This behaviour is also reproduced in simple models using one or two zones with spatially constant diffusion coefficients around the pulsars \citep{2020PhRvD.101j3035D}. We obtain a similar behaviour in our halo-profiles (see Figure \ref{fig:EmissionProfiles}) for those evolutionary stages during which CRs are still accelerated to the highest energies. However, near the cut-off energy the profiles transition to a more linear trend reflecting that those particles simply escape. This behaviour is more pronounced for electrons, on account of the synchrotron cooling of the high-energy electrons in the strong magnetic field inside the remnant.

\subsection{Reduction of the diffusion coefficient}

The escape of particles from the SNR and consequently the amplification of turbulence change the diffusion coefficient in the vicinity of the SNR \citep{2010ApJ...712L.153F,2011MNRAS.415.3434F}.
The escape of high-energy particles from their acceleration sites gained new attention with the detection of an extended halo around the Pulsars Geminga and PSR  B0656+14 \citep{2017Sci...358..911A}. 

Figure \ref{fig:Diffusion} shows the spatial variation of the diffusion coefficient relative to the assumed diffusion coefficient in the ISM, which is set to 10\% of the conventional Galactic diffusion coefficient \citep{2011ApJ...729..106T}. This choice was made to keep the time-step in our simulations reasonably large, otherwise the simulation of very old SNRs would be impossible. 

For both electrons and ions the reduction of the diffusion coefficient is strongest after $5,000\,$years ($10$\%-level) and $2,000\,$years ($1$\%-level), around the time of, or shortly after, the transition from free expansion to the Sedov-Taylor phase, when the maximum energy of the accelerated particles starts to decrease. Then there are too few freshly accelerated particles at high energy to sustain the level of turbulence, and consequently the diffusion coefficient starts to increase. The main damping mechanism of turbulence is cascading which, in contrast to other studies in the context of Pulsars \citep{2018PhRvD..98f3017E}, we treat not as a simple loss term but account for the energy transfer to smaller scales. Consequently, the distributions in Figure \ref{fig:Diffusion} show an energy-dependent time-evolution. The spatial extent is similar though, because the suppression of the diffusion coefficient for $100\,$GeV-particles is governed mainly by down-cascading from larger scales, i.e. by turbulence driven by particles at higher energy. Turbulence driving by the low-energy particles is only important very close to the shock.

To better understand the energy-dependent time evolution of upstream diffusion after the end of the free-expansion phase, it is instructive to closely inspect the cascading time \citep{Schlickeiser.2002a},
\begin{align}
    \tau(k) &= \frac{1}{v_{A}k}\sqrt{\frac{2U_B}{E_w}},\label{eq:DampingTime}
\end{align}
which depends on the wave-number, $k$, the spectral-energy density of the turbulent field, $E_{w}$, the Alfv\'en speed, $v_{A}$, and the energy density in the large-scale magnetic field, $U_{B}$. At all wave-numbers, except at $k_\text{min}$ resonant with the most energetic particles, turbulence energy is cascaded from large scales (small $k$) to small scales (large $k$). So, in a quasi-steady state, the level of turbulence at small scales is sustained by a continuous influx from larger scales. 

At an age of $4,000\,$ years, the cascading times for turbulence resonant with $3\,$TeV ($100\,$GeV) particles $5\,$pc ahead of the shock is about $500\,$years ($30\,$years). It takes $4,300$ years for the shock to arrive in this region, suggesting that cascading is efficient. However, the full spectral transport of turbulence is in many cases slower than the cascading time suggests, implying that the variation in the level of turbulence arises from a changing CR density gradient. As the freshly accelerated particles become fewer and less energetic with time, particles escaping from deep downstream become more important for turbulence driving.

During the initial growth of turbulence in the precursor, for the two particle energies shown in Figure \ref{fig:Diffusion} the diffusion coefficient shows the same trend, until the supply of freshly accelerated $3\,$TeV-particles is exhausted. However, the precursor scale of $100\,$GeV-particles is significantly smaller than for $3\,$TeV-particles, because the turbulence scattering the lower-energy particles results from cascading of modes resonant with the more energetic particles. Later, the escape of particles, that were trapped in the interior of the remnant, becomes the main driver of turbulence, and in fact the maximum extent of the turbulence precursor, and hence the reduced diffusion coefficient, is reached well after the beginning of the Sedov-phase. 

It is important to emphasize again that the diffusion coefficient for low-energy particles is determined by the turbulence that the high-energy particles provide. This has consequences beyond Supernova remnants, as for example low-energy particles in the halos of PWNs will reside closer to their acceleration sites than is suggested, if one treats cascading as a simple damping mechanism \citep{2018PhRvD..98f3017E}.

\subsection{Detectability}\label{sec:detectability}

\begin{figure}[h]
\includegraphics[width=0.485\textwidth]{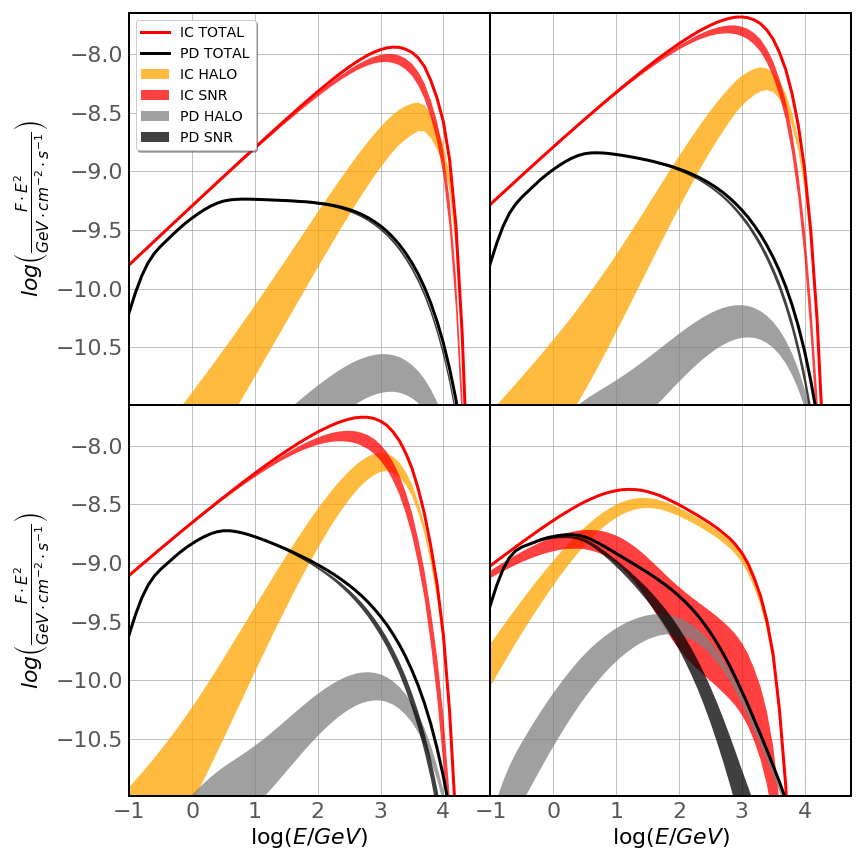}   
\caption{Comparison of emission spectra for IC (red) and PD (black) emission. Emission from the SNR is filled-black and filled-red for PD and IC-emission respectively. The halo-emission  is filled-orange and filled-gray for IC and PD-emission respectively at 300, 1000, 2,000 and 10,000yrs (from top left to bottom right). The spectra including (and excluding) the projection effect constitute the upper (lower) boundaries of the SNR emission and the lower (upper) limit of the area for emission from the halo.} 
\label{fig:Detectability}
\end{figure}

Figure \ref{fig:Detectability} shows the emission spectra expected from the SNR itself and the halo at different times. We assumed a generic distance of $1\,$kpc to the remnant. The contribution from the halo is shown twice, once as volume-integrated emission from $r > r_{sh}$, and once only the component that in projection appears to come from \mpo{beyond the projected radius} of the SNR. \mpo{We model three-dimensional objects, but we observe them in their 2D appearance in the sky. Thus, part of the emission from the halo will in projection appear to come from the SNR-interior, as the point of origin is in front of or behind the remnant}.

The evolution of the spectral energy distributions (SEDs) for PD-emission clearly shows the mechanism for the spectral softening: as highly energetic protons escape from the inside of the remnant, the low-energy particles remain inside, leading to a soft spectrum \citep{2020A&A...634A..59B}. The high-energy particles outside the SNR lack target material for the production of significant emission, and hence hadronic halo emission is unlikely detectable with current-generation gamma-ray observatories, even if the emission from the remnant and the halo could be disentangled. 

In case of the IC-emission, cooling becomes relevant after about $1,000\,$years, and the spectra from the interior of the remnant are modified by it. The halo emission from IC scattering is always much brighter than the PD component. Overall, the flux from the halo is only 20\% to 30\% of that of the SNR itself, making a detection possible only for the brightest known Galactic SNRs. As low-energy particles reside closer to the shock, they contribute little to the projected halo emission whose spectrum is consequently harder than the overall emission from the remnant. 

Figure \ref{fig:Detectability} demonstrates that leptonic emission detected from an SNR by current-generation gamma-ray observatories probably includes emission from the halo. Most of the halo emission is produced close to the remnant, and in projection it is very difficult to observationally separate it from emission from the interior. It will probably require the tenfold higher sensitivity of the future Cerenkov Telescope Array (CTA) to directly detect halo emission from the brightest TeV-remnants. What may have been seen already in RXJ1713.7-3946 is that the leptonic halo causes larger extent of the very-high gamma-ray emission compared to that of the X-ray emission. 

A special case is the remnant SN 1006 that has a bipolar morphology of non-thermal emission \citep{1981MNRAS.194..569P,2010A&A...516A..62A}, possibly originating from the variation of shock obliquity in a dominant large-scale magnetic field \citep{2003A&A...409..563V}. Regions with a quasi-perpendicular magnetic field are associated with inefficient ion acceleration \citep{2014ApJ...783...91C}. The apparent alignment of the bipolar morphology with the plane of the sky means that the emission from SN 1006 is much less distorted by projection effects than are more spherical remnants. If so, the contribution from the halo to the high-energy gamma-ray emission could be a factor of a few higher than that from other remnants of similar age. A direct measurement of the halo emission might still be difficult due to the limited spatial resolution of current-generation \rb{Imaging Air Cerenkov telescopes (IACTs)} and the low flux from SN 1006. In any case, a large fraction of the most energetic electrons resides in the upstream region (or halo). Due to the lower magnetic-field strength in that region, they contribute relatively little to the X-ray emission of SN 1006 (see also section \ref{sec:synchrotron}) and a lot to the gamma-ray emission. Figure \ref{fig:Detectability} indicates that this could lead to a spectral discrepancy between X-rays and high-energy gamma rays that is difficult to reproduce in a one-zone model. In the past, such a finding for SN 1006 was interpreted as effect of non-linear shock modification \citep{2010A&A...516A..62A}. We suggest that this discrepancy is simply a natural result of halo emission.

\subsection{Spectral index distribution}

The angular resolution of gamma-ray observatories is key to a study of spatial variations of the gamma-ray spectra \citep{2018A&A...612A...6H, 2018A&A...612A...7H, 2015ICRC...34..875H}. So far, no significant variation across the remnant has been detected.

\begin{figure}[h]
\includegraphics[width=0.49\textwidth]{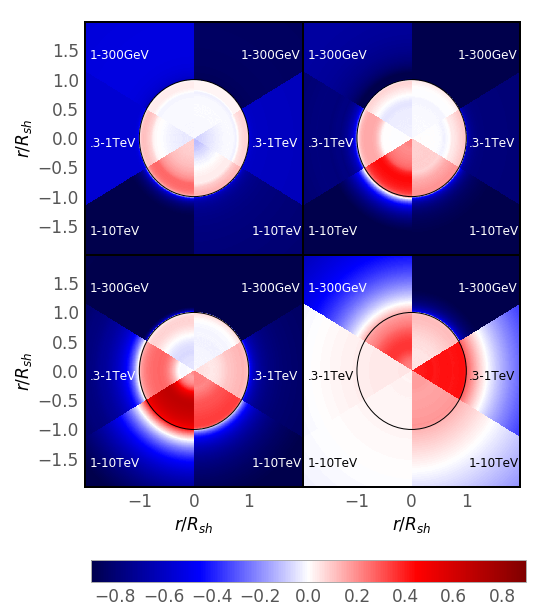}
\caption{Maps of spectral-index variation in three energy bands at 300, 1000, 2,000 and 10,000yrs (from top left to bottom right), relative to the average spectral index of the remnant (including the halo). Left hemispheres are inverse-Compton emission and right hemispheres Pion-decay emission.} 
\label{fig:IndexMaps}
\end{figure}
The spectrum of emission from outside the remnant is typically harder than that from the interior. Figure \ref{fig:IndexMaps} illustrates spectral index variations across the remnant at different ages. 
For emission from the interior of the remnant, the deviations are moderate\rb{. Depending on the energy-band, $\vert\Delta s\vert$ can reach values of up to $0.8$, and in general the emission \mpo{from the interior of} the remnant is softer than the average spectral index from the remnant. However, $\vert\Delta s\vert\lesssim 0.2$ is usually fulfilled for all but the highest energies of IC-emission. There, the mean spectral index \mpo{is close to that seen} upstream of the shock due to the large contribution of upstream-electrons to the gamma-ray flux. Still, the spectral index is quite uniform across the interior of the remnant, \mpo{whereas the} deviations of the spectral index of the halo emission are typically larger.} 

The detection of hadronic halo emission is unlikely given the current observational sensitivities. In the H.E.S.S. data for RXJ1713.7-3946 the $1$-$\sigma$ statistical uncertainty in the spatially-resolved spectral index maps is $\Delta s \approx \pm 0.1$ \citep{2018A&A...612A...6H}, likewise $\Delta s \approx \pm 0.15$ for Vela Junior \citep{2018A&A...612A...7H}. The systematic and the statistical uncertainties are comparable. In the VERITAS results for IC443, the spectral uncertainty ranges between $\Delta s =\pm 0.11$ and $\Delta s =\pm 0.42$ \citep{2015ICRC...34..875H}. Considering the limited spatial resolution and the poor statistics in some regions, a significant detection of spectral-index deviations is unlikely with current-generation IACTs.

The higher sensitivity of CTA should reduce the uncertainty, $\Delta s$, by roughly a factor of three, which should allow the detection of spectral-index variation with $3\sigma$ significance. We note that deviations from a spherical symmetry or asymmetric environments, such as density gradients in the ISM or fast motion of the progenitor \citep{2020MNRAS.493.3548M,2021MNRAS.502.5340M}, could lead to additional significant variations in the measured spectral index across the remnant.

\subsection{Synchrotron emission} \label{sec:synchrotron}
\begin{figure}[h]
\includegraphics[width=0.485\textwidth]{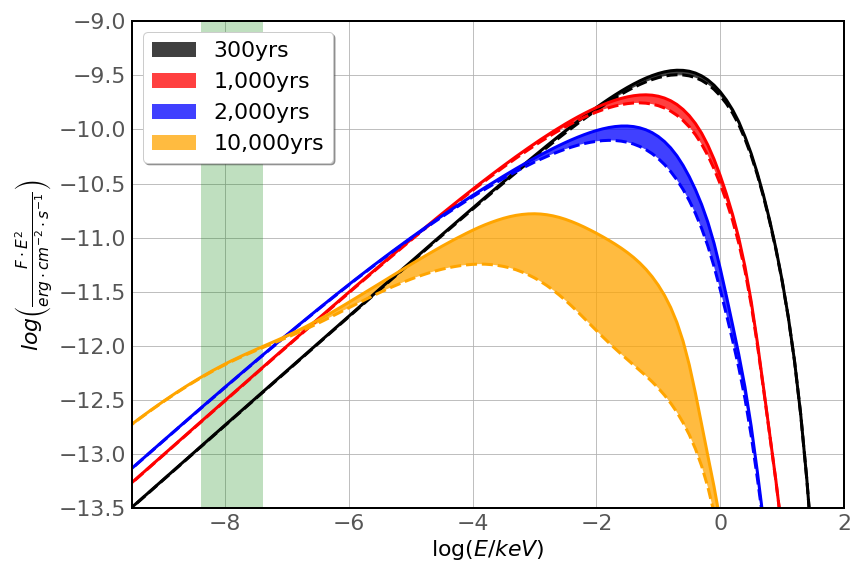}   
\caption{Synchrotron flux from the remnant at various times. The upper boundaries of the filled areas mark the total emission from the remnant while the lower boundaries indicate emission from downstream only. The green band indicates the $1-10\,$GHz range.}
\label{fig:SYCspectra}
\end{figure}

We evaluated the non-thermal synchrotron emission from the SNR over its lifetime. Figure \ref{fig:SYCspectra} shows the time-evolution of the total synchrotron flux, as well as the flux from the downstream region only.
Contrary to the IC-emission, there is little difference between the total and the downstream-only flux, except for very old remnants, because a strong magnetic field boosts the emission from the downstream. At later times, most of the high-energy electrons were able to escape the remnant and the magnetic field is not amplified any more, the halo emission starts to dominate.

The flux in the X-ray band decreases with time, because the magnetic-field strength decreases and cooling takes its toll in the electron spectra. Consequently, the remnant is brightest and best detectable during the first $2,000\,$yrs after the supernova - matching the census of Galactic SNRs \citep{2012A&ARv..20...49V}.

In contrast, the radio flux increases throughout \rb{the simulations}. Energy losses are not important, and the steady accumulation of GeV-scale particles compensates for the weakening magnetic field. Interestingly, the escape of electrons affects the spectral index in the radio band. Between 1 GHz and 30 GHz (Roughly $10^{-8}\,$keV to $10^{-7}\,$keV) the radio spectra are fairly soft after $10,000\,$yrs, with a spectral index $\alpha\approx 0.75$ ($S_\nu\propto \nu^{-\alpha}$). 

\begin{figure}[h]
\includegraphics[width=0.485\textwidth]{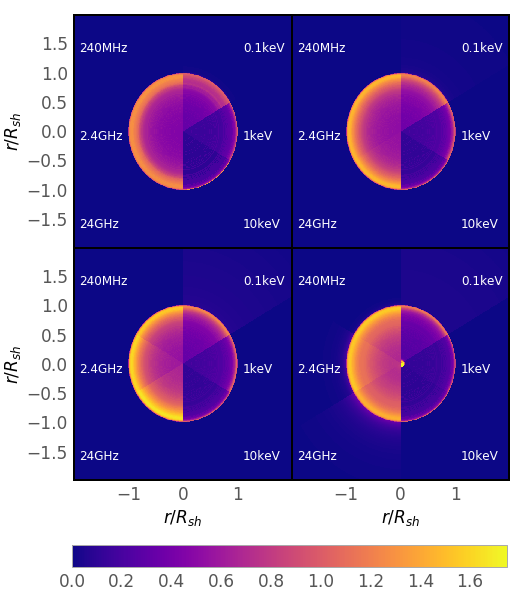}   
\caption{Maps of normalized synchrotron surface brightness of a Type-Ia SNR after 300, 1000, 2,000 and 10,000 yrs (from top left to bottom right).} 
\label{fig:SYCmaps}
\end{figure}

Figure \ref{fig:SYCmaps} shows the emission morphology at radio and X-ray energies for four stages of SNR evolution.
Throughout the SNR's lifetime the magnetic-field strength peaks right downstream of the forward shock. Initially, when the turbulent magnetic field exceeds the compressed large-scale field, cascading causes a decline toward the interior. Later, the trend persists due to the flow structure in the downstream of the shock. Consequently, the synchrotron emissivity peaks immediately downstream to the shock-location.

The X-ray morphology consists of a thin shell throughout the entire lifetime \rb{even though a part of the soft X-ray emission originates from the halo after $2000\,$years and might be detectable for bright sources.} The cutoff of the synchrotron spectrum is in or slightly below the X-ray band, and so the X-ray output is very sensitive to the structure of the magnetic field \citep{2005ApJ...626L.101P}. A slight shift in the characteristic synchrotron frequency due to a a spatial variation of the magnetic field can impose a strong
variation in the synchrotron flux.

The radio emission shows the same shell-like structure. However, the spatial variation of the magnetic field has a moderate impact on the synchrotron emissivity, typically $\propto B^{1.5}$. The radio shells appear thicker as a consequence. Moreover, diffusion of electron towards the center of the remnant partially compensates the weak magnetic field in the interior. Thus, the remnant appears somewhat center-filled in the radio band at later stages. \rb{Additionally, 4\% of the radio-emission at 10,000 years originates from the halo. The magnetic field is dominated by the large-scale field at that time, \mpo{which boosts the radio-emission from a downstream electron by a factor of $6$ compared to that of an upstream electron. This means that $20$\% of the low-energy electrons reside upstream of the shock and are escaping the remnant already.}}

To explain the center-filled radio morphology of the remnant, \cite{2014A&A...561A.139S} argued that additional, inward-moving shocks are needed to almost uniformly fill the interior of the Vela SNR with radio-emitting electrons. Our findings suggest that the same can be achieved by diffusive transport of electrons, in particular when in later stages of evolution the diffusion coefficient inside and outside of the remnant increases. However, Vela is likely the result of a core-collapse SN and thus additional factors need to be taken into account. Core-collapse SNRs initially evolve in the freely expanding wind of the progenitor star, in which the magnetic field will follow a $B\propto1/r$ dependence. Beyond the wind-termination shock, the magnetic-field strength is almost constant out to the contact discontinuity (\cite{2018A&A...618A.155S} and Sushch et al. (in preparation)), which strongly affects the X-ray and radio maps and likely enhances the trend towards a center-filled radio morphology.

\section{Conclusions}
We performed numerical simulations of particle acceleration in SNRs, solving time-dependent transport equations of CRs and magnetic turbulence in the test-particle limit alongside the standard gas-dynamical equations for Type-Ia SNRs. We derived the CR diffusion coefficient from the spectrum of magnetic turbulence, that evolves through driving by the resonant streaming instability as well as cascading and wave damping.

The gamma-ray morphology of SNRs depends on the radiation mechanism. Leptonic gamma-ray emission tends to produce a center-filled appearance with increasing remnant age, whereas hadronic gamma-ray emitters retain a shell-like morphology, on account of the distribution of target material for p-p interactions.

We showed that a gamma-ray halo has to be expected for leptonic emitters, despite the projection effects that place part of the halo emission inside the projected SNR radius. Whereas a firm detection of halo emission is in sight for the leptonic case, the halo component of hadronic emission likely remains undetectable, unless additional target material is present in the vicinity of the SNR. 

Our simulations show that the most energetic particles tend to reside outside the SNR. The relatively weak magnetic field in the upstream region leads to a low intensity of X-ray emission from these high-energetic electrons, making the X-ray spectrum appear softer than expected for a uniform magnetic field. It is likely that the observed spectral between X-ray and gamma-ray emission from SN 1006 is based on this effect instead of non-linear shock-modification.

Upstream of the forward shock, the diffusion coefficient is reduced, confining the accelerated particles close to the SNR. There is a strong spatial and temporal evolution of the diffusion coefficient in the entire precursor. The diffusion coefficient of low-energy particles is determined by cascading of turbulence downward from larger scales. In contrast to PWNs, for which neither the abundance of relativistic ions nor the time profile of particle acceleration are known, for SNRs we can self-consistently determine the injection spectra of CRs that will create the region of reduced diffusion around SNRs.

An investigation of the distribution of gamma-ray spectral indices across the interior of SNRs showed that no deviation beyond a $\vert\delta s\vert\approx0.2$ has to be expected for the known and resolvable Galactic SNRs. Detecting this small deviations is beyond the capabilities of current-generation IACTs but might be possible with CTA. Strong deviations from spherical symmetry can further enhance the detectability.

Initially the remnant emits non-thermal X-rays above $0.1\,$keV but cooling and a decreasing magnetic field reduce the X-ray luminosity and make a detection unlikely after a few thousand years. The radio flux from the remnant increases throughout its lifetime, and the radio morphology evolves from shell-like to a more center-filled appearance.

\section*{Acknowledgements}
Robert Brose acknowledges funding from an Irish Research Council Starting Laureate Award (IRCLA/2017/83).

\bibliographystyle{aa}
\bibliography{References}

\begin{thebibliography}{57}
\expandafter\ifx\csname natexlab\endcsname\relax\def\natexlab#1{#1}\fi

\bibitem[{{Abeysekara} {et~al.}(2017{\natexlab{a}}){Abeysekara}, {Albert},
  {Alfaro}, {Alvarez}, {{\'A}lvarez}, {Arceo}, {Arteaga-Vel{\'a}zquez}, {Avila
  Rojas}, {Ayala Solares}, {Barber}, {Bautista-Elivar}, {Becerril},
  {Belmont-Moreno}, {BenZvi}, {Berley}, {Bernal}, {Braun}, {Brisbois},
  {Caballero-Mora}, {Capistr{\'a}n}, {Carrami{\~n}ana}, {Casanova}, {Castillo},
  {Cotti}, {Cotzomi}, {Couti{\~n}o de Le{\'o}n}, {De Le{\'o}n}, {De la Fuente},
  {Dingus}, {DuVernois}, {D{\'\i}az-V{\'e}lez}, {Ellsworth}, {Engel},
  {Enr{\'\i}quez-Rivera}, {Fiorino}, {Fraija}, {Garc{\'\i}a-Gonz{\'a}lez},
  {Garfias}, {Gerhardt}, {Gonz{\'a}lez Mu{\~n}oz}, {Gonz{\'a}lez}, {Goodman},
  {Hampel-Arias}, {Harding}, {Hern{\'a}ndez}, {Hern{\'a}ndez-Almada}, {Hinton},
  {Hona}, {Hui}, {H{\"u}ntemeyer}, {Iriarte}, {Jardin-Blicq}, {Joshi},
  {Kaufmann}, {Kieda}, {Lara}, {Lauer}, {Lee}, {Lennarz}, {Vargas},
  {Linnemann}, {Longinotti}, {Luis Raya}, {Luna-Garc{\'\i}a}, {L{\'o}pez-Coto},
  {Malone}, {Marinelli}, {Martinez}, {Martinez-Castellanos},
  {Mart{\'\i}nez-Castro}, {Mart{\'\i}nez-Huerta}, {Matthews}, {Mirand
  a-Romagnoli}, {Moreno}, {Mostaf{\'a}}, {Nellen}, {Newbold}, {Nisa},
  {Noriega-Papaqui}, {Pelayo}, {Pretz}, {P{\'e}rez-P{\'e}rez}, {Ren}, {Rho},
  {Rivi{\`e}re}, {Rosa-Gonz{\'a}lez}, {Rosenberg}, {Ruiz-Velasco}, {Salazar},
  {Salesa Greus}, {Sand oval}, {Schneider}, {Schoorlemmer}, {Sinnis}, {Smith},
  {Springer}, {Surajbali}, {Taboada}, {Tibolla}, {Tollefson}, {Torres},
  {Ukwatta}, {Vianello}, {Weisgarber}, {Westerhoff}, {Wisher}, {Wood},
  {Yapici}, {Yodh}, {Younk}, {Zepeda}, {Zhou}, {Guo}, {Hahn}, {Li}, \&
  {Zhang}}]{2017Sci...358..911A}
{Abeysekara}, A.~U., {Albert}, A., {Alfaro}, R., {et~al.} 2017{\natexlab{a}},
  Science, 358, 911

\bibitem[{{Abeysekara} {et~al.}(2017{\natexlab{b}}){Abeysekara}, {Albert},
  {Alfaro}, {Alvarez}, {{\'A}lvarez}, {Arceo}, {Arteaga-Vel{\'a}zquez}, {Ayala
  Solares}, {Barber}, {Baughman}, {Bautista-Elivar}, {Becerra Gonzalez},
  {Becerril}, {Belmont-Moreno}, {BenZvi}, {Berley}, {Bernal}, {Braun},
  {Brisbois}, {Caballero-Mora}, {Capistr{\'a}n}, {Carrami{\~n}ana}, {Casanova},
  {Castillo}, {Cotti}, {Cotzomi}, {Couti{\~n}o de Le{\'o}n}, {de la Fuente},
  {De Le{\'o}n}, {Diaz Hernandez}, {Dingus}, {DuVernois},
  {D{\'\i}az-V{\'e}lez}, {Ellsworth}, {Engel}, {Fiorino}, {Fraija},
  {Garc{\'\i}a-Gonz{\'a}lez}, {Garfias}, {Gerhardt}, {Gonz{\'a}lez Mu{\~n}oz},
  {Gonz{\'a}lez}, {Goodman}, {Hampel-Arias}, {Harding}, {Hernand ez},
  {Hernandez-Almada}, {Hinton}, {Hui}, {H{\"u}ntemeyer}, {Iriarte},
  {Jardin-Blicq}, {Joshi}, {Kaufmann}, {Kieda}, {Lara}, {Lauer}, {Lee},
  {Lennarz}, {Le{\'o}n Vargas}, {Linnemann}, {Longinotti}, {Raya},
  {Luna-Garc{\'\i}a}, {L{\'o}pez-Coto}, {Malone}, {Marinelli}, {Martinez},
  {Martinez-Castellanos}, {Mart{\'\i}nez-Castro}, {Mart{\'\i}nez-Huerta},
  {Matthews}, {Mirand a-Romagnoli}, {Moreno}, {Mostaf{\'a}}, {Nellen},
  {Newbold}, {Nisa}, {Noriega-Papaqui}, {Pelayo}, {Pretz},
  {P{\'e}rez-P{\'e}rez}, {Ren}, {Rho}, {Rivi{\`e}re}, {Rosa-Gonz{\'a}lez},
  {Rosenberg}, {Ruiz-Velasco}, {Salazar}, {Salesa Greus}, {Sand oval},
  {Schneider}, {Schoorlemmer}, {Sinnis}, {Smith}, {Springer}, {Surajbali},
  {Taboada}, {Tibolla}, {Tollefson}, {Torres}, {Ukwatta}, {Vianello},
  {Villase{\~n}or}, {Weisgarber}, {Westerhoff}, {Wisher}, {Wood}, {Yapici},
  {Younk}, {Zepeda}, \& {Zhou}}]{2017ApJ...843...40A}
{Abeysekara}, A.~U., {Albert}, A., {Alfaro}, R., {et~al.} 2017{\natexlab{b}},
  \apj, 843, 40

\bibitem[{{Acero} {et~al.}(2010){Acero}, {Aharonian}, {Akhperjanian}, {Anton},
  {Barres de Almeida}, {Bazer-Bachi}, {Becherini}, {Behera}, {Beilicke},
  {Bernl{\"o}hr}, {Bochow}, {Boisson}, {Bolmont}, {Borrel}, {Brucker}, {Brun},
  {Brun}, {B{\"u}hler}, {Bulik}, {B{\"u}sching}, {Boutelier}, {Chadwick},
  {Charbonnier}, {Chaves}, {Cheesebrough}, {Conrad}, {Chounet}, {Clapson},
  {Coignet}, {Dalton}, {Daniel}, {Davids}, {Degrange}, {Deil}, {Dickinson},
  {Djannati-Ata{\"\i}}, {Domainko}, {O'C. Drury}, {Dubois}, {Dubus}, {Dyks},
  {Dyrda}, {Egberts}, {Eger}, {Espigat}, {Fallon}, {Farnier}, {Fegan},
  {Feinstein}, {Fiasson}, {F{\"o}rster}, {Fontaine}, {F{\"u}{\ss}ling},
  {Gabici}, {Gallant}, {G{\'e}rard}, {Gerbig}, {Giebels}, {Glicenstein},
  {Gl{\"u}ck}, {Goret}, {G{\"o}ring}, {Hauser}, {Hauser}, {Heinz},
  {Heinzelmann}, {Henri}, {Hermann}, {Hinton}, {Hoffmann}, {Hofmann},
  {Hofverberg}, {Holleran}, {Hoppe}, {Horns}, {Jacholkowska}, {de Jager},
  {Jahn}, {Jung}, {Katarzy{\'n}ski}, {Katz}, {Kaufmann}, {Kerschhaggl},
  {Khangulyan}, {Kh{\'e}lifi}, {Keogh}, {Klochkov}, {Klu{\'z}niak}, {Kneiske},
  {Komin}, {Kosack}, {Kossakowski}, {Lamanna}, {Lemoine-Goumard}, {Lenain},
  {Lohse}, {Marandon}, {Marcowith}, {Masbou}, {Maurin}, {McComb}, {Medina},
  {M{\'e}hault}, {Moderski}, {Moulin}, {Naumann-Godo}, {de Naurois}, {Nedbal},
  {Nekrassov}, {Nicholas}, {Niemiec}, {Nolan}, {Ohm}, {Olive}, {de O{\~n}a
  Wilhelmi}, {Orford}, {Ostrowski}, {Panter}, {Paz Arribas}, {Pedaletti},
  {Pelletier}, {Petrucci}, {Pita}, {P{\"u}hlhofer}, {Punch}, {Quirrenbach},
  {Raubenheimer}, {Raue}, {Rayner}, {Reimer}, {Renaud}, {de Los Reyes},
  {Rieger}, {Ripken}, {Rob}, {Rosier-Lees}, {Rowell}, {Rudak}, {Rulten},
  {Ruppel}, {Ryde}, {Sahakian}, {Santangelo}, {Schlickeiser}, {Sch{\"o}ck},
  {Sch{\"o}nwald}, {Schwanke}, {Schwarzburg}, {Schwemmer}, {Shalchi}, {Sushch},
  {Sikora}, {Skilton}, {Sol}, {Stawarz}, {Steenkamp}, {Stegmann}, {Stinzing},
  {Superina}, {Szostek}, {Tam}, {Tavernet}, {Terrier}, {Tibolla}, {Tluczykont},
  {van Eldik}, {Vasileiadis}, {Venter}, {Venter}, {Vialle}, {Vincent}, {Vink},
  {Vivier}, {V{\"o}lk}, {Volpe}, {Vorobiov}, {Wagner}, {Ward}, {Zdziarski},
  {Zech}, \& {H.~E.~S.~S. Collaboration}}]{2010A&A...516A..62A}
{Acero}, F., {Aharonian}, F., {Akhperjanian}, A.~G., {et~al.} 2010, \aap, 516,
  A62

\bibitem[{{Adriani} {et~al.}(2009){Adriani}, {Barbarino}, {Bazilevskaya},
  {Bellotti}, {Boezio}, {Bogomolov}, {Bonechi}, {Bongi}, {Bonvicini}, {Bottai},
  {Bruno}, {Cafagna}, {Campana}, {Carlson}, {Casolino}, {Castellini}, {de
  Pascale}, {de Rosa}, {de Simone}, {di Felice}, {Galper}, {Grishantseva},
  {Hofverberg}, {Koldashov}, {Krutkov}, {Kvashnin}, {Leonov}, {Malvezzi},
  {Marcelli}, {Menn}, {Mikhailov}, {Mocchiutti}, {Orsi}, {Osteria}, {Papini},
  {Pearce}, {Picozza}, {Ricci}, {Ricciarini}, {Simon}, {Sparvoli},
  {Spillantini}, {Stozhkov}, {Vacchi}, {Vannuccini}, {Vasilyev}, {Voronov},
  {Yurkin}, {Zampa}, {Zampa}, \& {Zverev}}]{2009Natur.458..607A}
{Adriani}, O., {Barbarino}, G.~C., {Bazilevskaya}, G.~A., {et~al.} 2009, \nat,
  458, 607

\bibitem[{{Bell}(1978)}]{1978MNRAS.182..147B}
{Bell}, A.~R. 1978, \mnras, 182, 147

\bibitem[{{Bell}(2004)}]{2004MNRAS.353..550B}
{Bell}, A.~R. 2004, \mnras, 353, 550

\bibitem[{{Beshley} \& {Petruk}(2012)}]{2012MNRAS.419.1421B}
{Beshley}, V. \& {Petruk}, O. 2012, \mnras, 419, 1421

\bibitem[{{Blasi} {et~al.}(2005){Blasi}, {Gabici}, \& {Vannoni}}]{Blasi.2005a}
{Blasi}, P., {Gabici}, S., \& {Vannoni}, G. 2005, \mnras, 361, 907

\bibitem[{{Bohdan} {et~al.}(2019){Bohdan}, {Niemiec}, {Pohl}, {Matsumoto},
  {Amano}, \& {Hoshino}}]{2019ApJ...878....5B}
{Bohdan}, A., {Niemiec}, J., {Pohl}, M., {et~al.} 2019, \apj, 878, 5

\bibitem[{{Brose} {et~al.}(2020){Brose}, {Pohl}, {Sushch}, {Petruk}, \&
  {Kuzyo}}]{2020A&A...634A..59B}
{Brose}, R., {Pohl}, M., {Sushch}, I., {Petruk}, O., \& {Kuzyo}, T. 2020, \aap,
  634, A59

\bibitem[{{Brose} {et~al.}(2016){Brose}, {Telezhinsky}, \&
  {Pohl}}]{2016A&A...593A..20B}
{Brose}, R., {Telezhinsky}, I., \& {Pohl}, M. 2016, \aap, 593, A20

\bibitem[{{Caprioli} \& {Spitkovsky}(2014)}]{2014ApJ...783...91C}
{Caprioli}, D. \& {Spitkovsky}, A. 2014, \apj, 783, 91

\bibitem[{{Celli} {et~al.}(2019){Celli}, {Morlino}, {Gabici}, \&
  {Aharonian}}]{2019MNRAS.490.4317C}
{Celli}, S., {Morlino}, G., {Gabici}, S., \& {Aharonian}, F.~A. 2019, \mnras,
  490, 4317

\bibitem[{{Chang} {et~al.}(2008){Chang}, {Adams}, {Ahn}, {Bashindzhagyan},
  {Christl}, {Ganel}, {Guzik}, {Isbert}, {Kim}, {Kuznetsov}, {Panasyuk},
  {Panov}, {Schmidt}, {Seo}, {Sokolskaya}, {Watts}, {Wefel}, {Wu}, \&
  {Zatsepin}}]{2008Natur.456..362C}
{Chang}, J., {Adams}, J.~H., {Ahn}, H.~S., {et~al.} 2008, \nat, 456, 362

\bibitem[{{Di Mauro} {et~al.}(2020){Di Mauro}, {Manconi}, \&
  {Donato}}]{2020PhRvD.101j3035D}
{Di Mauro}, M., {Manconi}, S., \& {Donato}, F. 2020, \prd, 101, 103035

\bibitem[{{Dwarkadas} \& {Chevalier}(1998)}]{1998ApJ...497..807D}
{Dwarkadas}, V.~V. \& {Chevalier}, R.~A. 1998, \apj, 497, 807

\bibitem[{{Evoli} {et~al.}(2018){Evoli}, {Linden}, \&
  {Morlino}}]{2018PhRvD..98f3017E}
{Evoli}, C., {Linden}, T., \& {Morlino}, G. 2018, \prd, 98, 063017

\bibitem[{{Fujita} {et~al.}(2010){Fujita}, {Ohira}, \&
  {Takahara}}]{2010ApJ...712L.153F}
{Fujita}, Y., {Ohira}, Y., \& {Takahara}, F. 2010, \apjl, 712, L153

\bibitem[{{Fujita} {et~al.}(2011){Fujita}, {Takahara}, {Ohira}, \&
  {Iwasaki}}]{2011MNRAS.415.3434F}
{Fujita}, Y., {Takahara}, F., {Ohira}, Y., \& {Iwasaki}, K. 2011, \mnras, 415,
  3434

\bibitem[{{Giacinti} {et~al.}(2020){Giacinti}, {Mitchell}, {L{\'o}pez-Coto},
  {Joshi}, {Parsons}, \& {Hinton}}]{2020A&A...636A.113G}
{Giacinti}, G., {Mitchell}, A.~M.~W., {L{\'o}pez-Coto}, R., {et~al.} 2020,
  \aap, 636, A113

\bibitem[{{H.~E.~S.~S. Collaboration} {et~al.}(2018{\natexlab{a}}){H.~E.~S.~S.
  Collaboration}, {Abdalla}, {Abramowski}, {Aharonian}, {Ait Benkhali},
  {Akhperjanian}, {Andersson}, {Ang{\"u}ner}, {Arrieta}, {Aubert}, {Backes},
  {Balzer}, {Barnard}, {Becherini}, {Becker Tjus}, {Berge}, {Bernhard},
  {Bernl{\"o}hr}, {Blackwell}, {B{\"o}ttcher}, {Boisson}, {Bolmont}, {Bordas},
  {Bregeon}, {Brun}, {Brun}, {Bryan}, {Bulik}, {Capasso}, {Carr}, {Casanova},
  {Cerruti}, {Chakraborty}, {Chalme-Calvet}, {Chaves}, {Chen}, {Chevalier},
  {Chr{\'e}tien}, {Colafrancesco}, {Cologna}, {Condon}, {Conrad}, {Cui},
  {Davids}, {Decock}, {Degrange}, {Deil}, {Devin}, {deWilt}, {Dirson},
  {Djannati-Ata{\"\i}}, {Domainko}, {Donath}, {Drury}, {Dubus}, {Dutson},
  {Dyks}, {Edwards}, {Egberts}, {Eger}, {Ernenwein}, {Eschbach}, {Farnier},
  {Fegan}, {Fernand es}, {Fiasson}, {Fontaine}, {F{\"o}rster}, {Fukuyama},
  {Funk}, {F{\"u}{\ss}ling}, {Gabici}, {Gajdus}, {Gallant}, {Garrigoux},
  {Giavitto}, {Giebels}, {Glicenstein}, {Gottschall}, {Goyal}, {Grondin},
  {Hadasch}, {Hahn}, {Haupt}, {Hawkes}, {Heinzelmann}, {Henri}, {Hermann},
  {Hervet}, {Hinton}, {Hofmann}, {Hoischen}, {Holler}, {Horns}, {Ivascenko},
  {Jacholkowska}, {Jamrozy}, {Janiak}, {Jankowsky}, {Jankowsky}, {Jingo},
  {Jogler}, {Jouvin}, {Jung-Richardt}, {Kastendieck}, {Katarzy{\'n}ski},
  {Katz}, {Kerszberg}, {Kh{\'e}lifi}, {Kieffer}, {King}, {Klepser}, {Klochkov},
  {Klu{\'z}niak}, {Kolitzus}, {Komin}, {Kosack}, {Krakau}, {Kraus}, {Krayzel},
  {Kr{\"u}ger}, {Laffon}, {Lamanna}, {Lau}, {Lees}, {Lefaucheur}, {Lefranc},
  {Lemi{\`e}re}, {Lemoine-Goumard}, {Lenain}, {Leser}, {Lohse}, {Lorentz},
  {Liu}, {L{\'o}pez-Coto}, {Lypova}, {Marandon}, {Marcowith}, {Mariaud},
  {Marx}, {Maurin}, {Maxted}, {Mayer}, {Meintjes}, {Meyer}, {Mitchell},
  {Moderski}, {Mohamed}, {Mohrmann}, {Mor{\r{a}}}, {Moulin}, {Murach}, {de
  Naurois}, {Niederwanger}, {Niemiec}, {Oakes}, {O'Brien}, {Odaka}, {{\"O}ttl},
  {Ohm}, {Ostrowski}, {Oya}, {Padovani}, {Panter}, {Parsons}, {Pekeur},
  {Pelletier}, {Perennes}, {Petrucci}, {Peyaud}, {Piel}, {Pita}, {Poon},
  {Prokhorov}, {Prokoph}, {P{\"u}hlhofer}, {Punch}, {Quirrenbach}, {Raab},
  {Reimer}, {Reimer}, {Renaud}, {de los Reyes}, {Rieger}, {Romoli},
  {Rosier-Lees}, {Rowell}, {Rudak}, {Rulten}, {Sahakian}, {Salek}, {Sanchez},
  {Santangelo}, {Sasaki}, {Schlickeiser}, {Sch{\"u}ssler}, {Schulz},
  {Schwanke}, {Schwemmer}, {Settimo}, {Seyffert}, {Shafi}, {Shilon}, {Simoni},
  {Sol}, {Spanier}, {Spengler}, {Spies}, {Stawarz}, {Steenkamp}, {Stegmann},
  {Stinzing}, {Stycz}, {Sushch}, {Takahashi}, {Tavernet}, {Tavernier},
  {Taylor}, {Terrier}, {Tibaldo}, {Tiziani}, {Tluczykont}, {Trichard}, {Tuffs},
  {Uchiyama}, {van der Walt}, {van Eldik}, {van Rensburg}, {van Soelen},
  {Vasileiadis}, {Veh}, {Venter}, {Viana}, {Vincent}, {Vink}, {Voisin},
  {V{\"o}lk}, {Volpe}, {Vuillaume}, {Wadiasingh}, {Wagner}, {Wagner}, {Wagner},
  {White}, {Wierzcholska}, {Willmann}, {W{\"o}rnlein}, {Wouters}, {Yang},
  {Zabalza}, {Zaborov}, {Zacharias}, {Zdziarski}, {Zech}, {Zefi}, {Ziegler}, \&
  {{\.Z}ywucka}}]{2018A&A...612A...6H}
{H.~E.~S.~S. Collaboration}, {Abdalla}, H., {Abramowski}, A., {et~al.}
  2018{\natexlab{a}}, \aap, 612, A6

\bibitem[{{H.~E.~S.~S. Collaboration} {et~al.}(2018{\natexlab{b}}){H.~E.~S.~S.
  Collaboration}, {Abdalla}, {Abramowski}, {Aharonian}, {Ait Benkhali},
  {Akhperjanian}, {Ang{\"u}ner}, {Arakawa}, {Arrieta}, {Aubert}, {Backes},
  {Balzer}, {Barnard}, {Becherini}, {Becker Tjus}, {Berge}, {Bernhard},
  {Bernl{\"o}hr}, {Blackwell}, {B{\"o}ttcher}, {Boisson}, {Bolmont}, {Bordas},
  {Bregeon}, {Brun}, {Brun}, {Bryan}, {B{\"u}chele}, {Bulik}, {Capasso},
  {Carr}, {Casanova}, {Cerruti}, {Chakraborty}, {Chalme-Calvet}, {Chaves},
  {Chen}, {Chevalier}, {Chr{\'e}tien}, {Coffaro}, {Colafrancesco}, {Cologna},
  {Condon}, {Conrad}, {Cui}, {Davids}, {Decock}, {Degrange}, {Deil}, {Devin},
  {deWilt}, {Dirson}, {Djannati-Ata{\"\i}}, {Domainko}, {Donath}, {Drury},
  {Dutson}, {Dyks}, {Edwards}, {Egberts}, {Eger}, {Ernenwein}, {Eschbach},
  {Farnier}, {Fegan}, {Fernand es}, {Fiasson}, {Fontaine}, {F{\"o}rster},
  {Funk}, {F{\"u}{\ss}ling}, {Gabici}, {Gajdus}, {Gallant}, {Garrigoux},
  {Giavitto}, {Giebels}, {Glicenstein}, {Gottschall}, {Goyal}, {Grondin},
  {Hahn}, {Haupt}, {Hawkes}, {Heinzelmann}, {Henri}, {Hermann}, {Hervet},
  {Hinton}, {Hofmann}, {Hoischen}, {Holler}, {Horns}, {Ivascenko}, {Iwasaki},
  {Jacholkowska}, {Jamrozy}, {Janiak}, {Jankowsky}, {Jankowsky}, {Jingo},
  {Jogler}, {Jouvin}, {Jung-Richardt}, {Kastendieck}, {Katarzy{\'n}ski},
  {Katsuragawa}, {Katz}, {Kerszberg}, {Khangulyan}, {Kh{\'e}lifi}, {Kieffer},
  {King}, {Klepser}, {Klochkov}, {Klu{\'z}niak}, {Kolitzus}, {Komin}, {Krakau},
  {Kraus}, {Kr{\"u}ger}, {Laffon}, {Lamanna}, {Lau}, {Lees}, {Lefaucheur},
  {Lefranc}, {Lemi{\`e}re}, {Lemoine-Goumard}, {Lenain}, {Leser}, {Lohse},
  {Lorentz}, {Liu}, {L{\'o}pez-Coto}, {Lypova}, {Marandon}, {Marcowith},
  {Mariaud}, {Marx}, {Maurin}, {Maxted}, {Mayer}, {Meintjes}, {Meyer},
  {Mitchell}, {Moderski}, {Mohamed}, {Mohrmann}, {Mor{\r{a}}}, {Moulin},
  {Murach}, {Nakashima}, {de Naurois}, {Niederwanger}, {Niemiec}, {Oakes},
  {O'Brien}, {Odaka}, {{\"O}ttl}, {Ohm}, {Ostrowski}, {Oya}, {Padovani},
  {Panter}, {Parsons}, {Paz Arribas}, {Pekeur}, {Pelletier}, {Perennes},
  {Petrucci}, {Peyaud}, {Piel}, {Pita}, {Poon}, {Prokhorov}, {Prokoph},
  {P{\"u}hlhofer}, {Punch}, {Quirrenbach}, {Raab}, {Reimer}, {Reimer},
  {Renaud}, {de los Reyes}, {Richter}, {Rieger}, {Romoli}, {Rowell}, {Rudak},
  {Rulten}, {Sahakian}, {Saito}, {Salek}, {Sanchez}, {Santangelo}, {Sasaki},
  {Schlickeiser}, {Sch{\"u}ssler}, {Schulz}, {Schwanke}, {Schwemmer},
  {Seglar-Arroyo}, {Settimo}, {Seyffert}, {Shafi}, {Shilon}, {Simoni}, {Sol},
  {Spanier}, {Spengler}, {Spies}, {Stawarz}, {Steenkamp}, {Stegmann}, {Stycz},
  {Sushch}, {Takahashi}, {Tavernet}, {Tavernier}, {Taylor}, {Terrier},
  {Tibaldo}, {Tiziani}, {Tluczykont}, {Trichard}, {Tsuji}, {Tuffs}, {Uchiyama},
  {van der Walt}, {van Eldik}, {van Rensburg}, {van Soelen}, {Vasileiadis},
  {Veh}, {Venter}, {Viana}, {Vincent}, {Vink}, {Voisin}, {V{\"o}lk},
  {Vuillaume}, {Wadiasingh}, {Wagner}, {Wagner}, {Wagner}, {White},
  {Wierzcholska}, {Willmann}, {W{\"o}rnlein}, {Wouters}, {Yang}, {Zabalza},
  {Zaborov}, {Zacharias}, {Zanin}, {Zdziarski}, {Zech}, {Zefi}, {Ziegler}, \&
  {{\.Z}ywucka}}]{2018A&A...612A...7H}
{H.~E.~S.~S. Collaboration}, {Abdalla}, H., {Abramowski}, A., {et~al.}
  2018{\natexlab{b}}, \aap, 612, A7

\bibitem[{{Humensky} \& {VERITAS Collaboration}(2015)}]{2015ICRC...34..875H}
{Humensky}, B. \& {VERITAS Collaboration}. 2015, in International Cosmic Ray
  Conference, Vol.~34, 34th International Cosmic Ray Conference (ICRC2015), 875

\bibitem[{{Kobzar} {et~al.}(2017){Kobzar}, {Niemiec}, {Pohl}, \&
  {Bohdan}}]{2017MNRAS.469.4985K}
{Kobzar}, O., {Niemiec}, J., {Pohl}, M., \& {Bohdan}, A. 2017, \mnras, 469,
  4985

\bibitem[{{Li} {et~al.}(2018){Li}, {Zhou}, {Huang}, {Zhang}, {Qiao}, {Yu},
  {Ruan}, \& {He}}]{2018PhPl...25h2103L}
{Li}, R., {Zhou}, C.~T., {Huang}, T.~W., {et~al.} 2018, Physics of Plasmas, 25,
  082103

\bibitem[{{Lucek} \& {Bell}(2000)}]{2000MNRAS.314...65L}
{Lucek}, S.~G. \& {Bell}, A.~R. 2000, \mnras, 314, 65

\bibitem[{{Malkov}(1998)}]{1998PhRvE..58.4911M}
{Malkov}, M.~A. 1998, \pre, 58, 4911

\bibitem[{{Matsumoto} {et~al.}(2017){Matsumoto}, {Amano}, {Kato}, \&
  {Hoshino}}]{Matsumoto2017}
{Matsumoto}, Y., {Amano}, T., {Kato}, T.~N., \& {Hoshino}, M. 2017, Phys. Rev.
  Lett.

\bibitem[{{Meyer} {et~al.}(2020){Meyer}, {Petrov}, \&
  {Pohl}}]{2020MNRAS.493.3548M}
{Meyer}, D.~M.~A., {Petrov}, M., \& {Pohl}, M. 2020, \mnras, 493, 3548

\bibitem[{{Meyer} {et~al.}(2021){Meyer}, {Pohl}, {Petrov}, \&
  {Oskinova}}]{2021MNRAS.502.5340M}
{Meyer}, D.~M.~A., {Pohl}, M., {Petrov}, M., \& {Oskinova}, L. 2021, \mnras,
  502, 5340

\bibitem[{{Mignone} {et~al.}(2007){Mignone}, {Bodo}, {Massaglia}, {Matsakos},
  {Tesileanu}, {Zanni}, \& {Ferrari}}]{2007ApJS..170..228M}
{Mignone}, A., {Bodo}, G., {Massaglia}, S., {et~al.} 2007, \apjs, 170, 228

\bibitem[{{Nava} {et~al.}(2016){Nava}, {Gabici}, {Marcowith}, {Morlino}, \&
  {Ptuskin}}]{2016MNRAS.461.3552N}
{Nava}, L., {Gabici}, S., {Marcowith}, A., {Morlino}, G., \& {Ptuskin}, V.~S.
  2016, \mnras, 461, 3552

\bibitem[{{Niemiec} {et~al.}(2010){Niemiec}, {Pohl}, {Bret}, \&
  {Stroman}}]{2010ApJ...709.1148N}
{Niemiec}, J., {Pohl}, M., {Bret}, A., \& {Stroman}, T. 2010, \apj, 709, 1148

\bibitem[{{Ohira} {et~al.}(2010){Ohira}, {Murase}, \&
  {Yamazaki}}]{2010A&A...513A..17O}
{Ohira}, Y., {Murase}, K., \& {Yamazaki}, R. 2010, \aap, 513, A17

\bibitem[{{Pais} \& {Pfrommer}(2020)}]{2020MNRAS.498.5557P}
{Pais}, M. \& {Pfrommer}, C. 2020, \mnras, 498, 5557

\bibitem[{{Petruk} {et~al.}(2009){Petruk}, {Beshley}, {Bocchino}, \&
  {Orlando}}]{2009MNRAS.395.1467P}
{Petruk}, O., {Beshley}, V., {Bocchino}, F., \& {Orlando}, S. 2009, \mnras,
  395, 1467

\bibitem[{{Pohl} {et~al.}(2005){Pohl}, {Yan}, \&
  {Lazarian}}]{2005ApJ...626L.101P}
{Pohl}, M., {Yan}, H., \& {Lazarian}, A. 2005, \apjl, 626, L101

\bibitem[{{Principe} {et~al.}(2020){Principe}, {Mitchell}, {Caroff}, {Hinton},
  {Parsons}, \& {Funk}}]{2020arXiv200611177P}
{Principe}, G., {Mitchell}, A.~M.~W., {Caroff}, S., {et~al.} 2020, arXiv
  e-prints, arXiv:2006.11177

\bibitem[{{Profumo} {et~al.}(2018){Profumo}, {Reynoso-Cordova}, {Kaaz}, \&
  {Silverman}}]{2018PhRvD..97l3008P}
{Profumo}, S., {Reynoso-Cordova}, J., {Kaaz}, N., \& {Silverman}, M. 2018,
  \prd, 97, 123008

\bibitem[{{Ptuskin} \& {Zirakashvili}(2003)}]{2003A&A...403....1P}
{Ptuskin}, V.~S. \& {Zirakashvili}, V.~N. 2003, \aap, 403, 1

\bibitem[{{Pye} {et~al.}(1981){Pye}, {Punds}, {Rolf}, {Seward}, {Smith}, \&
  {Willingale}}]{1981MNRAS.194..569P}
{Pye}, J.~P., {Punds}, K.~A., {Rolf}, D.~P., {et~al.} 1981, \mnras, 194, 569

\bibitem[{{Reimer} {et~al.}(2006){Reimer}, {Pohl}, \&
  {Reimer}}]{2006ApJ...644.1118R}
{Reimer}, A., {Pohl}, M., \& {Reimer}, O. 2006, \apj, 644, 1118

\bibitem[{{Reville} \& {Bell}(2013)}]{2013MNRAS.430.2873R}
{Reville}, B. \& {Bell}, A.~R. 2013, \mnras, 430, 2873

\bibitem[{{Riquelme} \& {Spitkovsky}(2009)}]{2009ApJ...694..626R}
{Riquelme}, M.~A. \& {Spitkovsky}, A. 2009, \apj, 694, 626

\bibitem[{{Schlickeiser}(2002)}]{Schlickeiser.2002a}
{Schlickeiser}, R. 2002, {Cosmic Ray Astrophysics}

\bibitem[{{Skilling}(1975)}]{Skilling.1975a}
{Skilling}, J. 1975, MNRAS, 172, 557

\bibitem[{{Sushch} {et~al.}(2018){Sushch}, {Brose}, \&
  {Pohl}}]{2018A&A...618A.155S}
{Sushch}, I., {Brose}, R., \& {Pohl}, M. 2018, \aap, 618, A155

\bibitem[{{Sushch} \& {Hnatyk}(2014)}]{2014A&A...561A.139S}
{Sushch}, I. \& {Hnatyk}, B. 2014, \aap, 561, A139

\bibitem[{{Sutherland} \& {Dopita}(1993)}]{1993ApJS...88..253S}
{Sutherland}, R.~S. \& {Dopita}, M.~A. 1993, \apjs, 88, 253

\bibitem[{{Telezhinsky} {et~al.}(2013){Telezhinsky}, {Dwarkadas}, \&
  {Pohl}}]{2013A&A...552A.102T}
{Telezhinsky}, I., {Dwarkadas}, V.~V., \& {Pohl}, M. 2013, \aap, 552, A102

\bibitem[{{Trotta} {et~al.}(2011){Trotta}, {J{\'o}hannesson}, {Moskalenko},
  {Porter}, {Ruiz de Austri}, \& {Strong}}]{2011ApJ...729..106T}
{Trotta}, R., {J{\'o}hannesson}, G., {Moskalenko}, I.~V., {et~al.} 2011, \apj,
  729, 106

\bibitem[{{Vink}(2006)}]{2006ESASP.604..319V}
{Vink}, J. 2006, in ESA Special Publication, Vol. 604, The X-ray Universe 2005,
  ed. A.~{Wilson}, 319

\bibitem[{{Vink}(2012)}]{2012A&ARv..20...49V}
{Vink}, J. 2012, \aapr, 20, 49

\bibitem[{{V{\"o}lk} {et~al.}(2003){V{\"o}lk}, {Berezhko}, \&
  {Ksenofontov}}]{2003A&A...409..563V}
{V{\"o}lk}, H.~J., {Berezhko}, E.~G., \& {Ksenofontov}, L.~T. 2003, A\&A, 409,
  563

\bibitem[{{Yan} {et~al.}(2012){Yan}, {Lazarian}, \&
  {Schlickeiser}}]{2012ApJ...745..140Y}
{Yan}, H., {Lazarian}, A., \& {Schlickeiser}, R. 2012, \apj, 745, 140

\bibitem[{{Zeng} {et~al.}(2019){Zeng}, {Xin}, \& {Liu}}]{2019ApJ...874...50Z}
{Zeng}, H., {Xin}, Y., \& {Liu}, S. 2019, \apj, 874, 50

\bibitem[{{Zhou} \& {Matthaeus}(1990)}]{1990JGR....9514881Z}
{Zhou}, Y. \& {Matthaeus}, W.~H. 1990, \jgr, 95, 14881

\end{thebibliography}

\appendix 
\section{Electron injection derived from SN1006} \label{sec:InjSN1006}
The injection fraction, $\eta_i$, is a crucial parameter for our simulations. The normalization of the proton spectrum determines the gradient of the CR distribution in the upstream region and hence the growth rate of magnetic turbulence. At late times, $\eta_i$ can be understood as the number ratio of particles injected into shock acceleration and the total number of thermal particles in the system,
\begin{align}
    \eta_i &= \frac{\int_{E_i}^\infty N(E)\text{d} E}{\int \rho_0/m \text{d}V} \text{ , }
\end{align}
where $E_i$ is the injection energy, $N(E)$ the differential number density of CRs, and $\rho_0/m$ the number density of the ambient medium. If $\eta_i$ is known, equation (\ref{eq:Injection}) can be used to determine $\psi$ for the setup of the simulation.

$\eta_i$ can also be determined from SNR observations. We pointed out that the luminosity-age distribution of Galactic SNRs can be related to the total number of accelerated electrons in these remnants, if one considers the density profile for the Sedov solution. Then, the gamma-ray emission above $1\,$TeV is dominated by the IC channel \citep{2020A&A...634A..59B}. 

If the ambient density is very low, as it is for SN~1006, $\eta_i$ can also be inferred directly, because it is likely that all of the high-energy photons are produced via IC emission.
The analysis of the H.E.S.S.-collaboration revealed a total energy in electrons of $W_e=3.3\cdot 10^{47}\,$erg, an electron spectral-index of $s_e=2.1$, and a electron cutoff-energy of $E_c=10\,$TeV \citep{2010A&A...516A..62A}. These values can now be used to determine the total number of electrons present in SN 1006. We assume for simplicity that energetic electrons are present with constant spectral index between $E_0=10\,$MeV and $E_c=10\,$TeV.
\begin{align}
    W_e &= \int_{E_0}^{E_c} E\cdot N(E) \text{d}E\\
    N(E) &= N_0\cdot E^{-s} \text{ for $E_0< E < E_c$}\\
    \frac{W_e}{\int_{E_0}^{E_c}N(E)\text{d}E}&=\frac{E_c^{2-s}-E_0^{2-s}}{E_c^{1-s}-E_0^{1-s}}\frac{1-s}{2-s}=\alpha\\
    \int_{E_0}^{E_c}N(E)\text{d}E &= \frac{W_e}{\alpha}=6.4\cdot 10^{51}
\end{align}
The total number of thermal particles can be calculated by integrating the density of the ambient medium over the volume filled by the SNR. SN 1006 has an reported radius of $R=9.6\,$pc in a medium with $\rho_0/m=0.085\,$cm$^{-3}$.
\begin{align}
    \int\frac{\rho_0}{m}\text{d}V &= \frac{4}{3}\pi R^3\frac{\rho_0}{m}=9.26\cdot 10^{57} \text{ . }
\end{align}
This gives $\eta_i=6.9\cdot10^{-7}$ for electrons. A correction is in order, because not the entire shock surface of SN 1006 is efficiently accelerating particles. Estimating that only $~20\,$\% of SN 1006's surface are contributing to the high-energy emission, we find $\eta_i=3.5\cdot10^{-6}$, which corresponds to $\psi=4.205$. This is very close to the value $\psi=4.2$ that we used in our simulations. 

Experimental data suggests that for particles arriving at earth the electron-to-proton ratio is $K_{ep}=0.01$. We inject $1.2$ times more electrons than protons to account for a composition with $10\,$\% helium. Since electrons and protons are injected at the same $\psi$ but at different energies, this results in $K_{ep}\approx0.01$.

\end{document}